\documentstyle[preprint,epsfig,eqsecnum,aps,floats,tighten]{revtex}

\def\Missing#1#2{{\mbox{$#1\kern-0.57em\raise0.19ex\hbox{/}_{#2}$}}}

\def\vMissing#1#2{\ifmmode
            \vec{#1}\kern-0.57em\raise.19ex\hbox{/}_{#2}
         \else
            {{\mbox{$\vec{#1}\kern-0.57em\raise.19ex\hbox{/}_{#2}$}}}
         \fi}
\def\lsim{\mathrel{\rlap{\lower4pt\hbox{\hskip1pt$\sim$}}
    \raise1pt\hbox{$<$}}}        
\def\gsim{\mathrel{\rlap{\lower4pt\hbox{\hskip1pt$\sim$}}
    \raise1pt\hbox{$>$}}}

\def\D0{D\O }
\def\etal{\it et al}

\def\dsdpt{\mbox{$d\sigma^W/dp_T^W$}}

\def\simge{\mathrel{\rlap{\raise 0.53ex \hbox{$>$}}%
{\lower 0.53ex \hbox{$\sim$}}}}
\def\simle{\mathrel{\rlap{\raise 0.53ex \hbox{$<$}}%
{\lower 0.53ex \hbox{$\sim$}}}}

\def\ETmiss{\mbox{${\hbox{$E$\kern-0.5em\lower-.1ex\hbox{/}\kern+0.15em}}_T$ }}

\def\err#1#2#3 {{\it Erratum} {\bf#1},{\ #2} (19#3)}
\def\ib#1#2#3 {{\it ibid.} {\bf#1},{\ #2} (19#3)}
\def\nc#1#2#3 {Nuovo Cim. {\bf#1} ,#2(19#3)}
\def\nim#1#2#3 {Nucl. Instr. Meth. {\bf#1},{\ #2} (19#3)}
\def\np#1#2#3 {Nucl. Phys. {\bf#1},{\ #2} (19#3)}
\def\pl#1#2#3 {Phys. Lett. {\bf#1},{\ #2} (19#3)}
\def\prev#1#2#3 {Phys. Rev. {\bf#1},{\ #2} (19#3)}
\def\prl#1#2#3 {Phys. Rev. Lett. {\bf#1},{\ #2} (19#3)}
\def\rmp#1#2#3 {Rev. Mod. Phys. {\bf#1},{\ #2} (19#3)}
\def\zp#1#2#3 {Zeit. Phys. {\bf#1},{\ #2} (19#3)}     
\begin{document}

%\preprint{FERMILAB-Pub-97/109-E}
%
% ======> Title of the paper goes here <====================
%
\title{Measurement of the ratio of the differential cross sections
for $W$ and $Z$ boson production as a function of transverse momentum
       in {\mbox{$p\bar p$}}\ collisions at 
       {\mbox{$\sqrt{s}$ =\ 1.8\ TeV}} }
                                                    
\author{\centerline{The D\O\ Collaboration
  \thanks{Submitted to the {\it International Europhysics Conference
        on High Energy Physics},
	\hfill\break
	July 12-18, 2001, Budapest, Hungary,
        \hfill\break 
	and  {\it XX International Symposium on Lepton and Photon Interactions at High Energies}
	\hfill\break
        July 23 -- 28, 2001, Rome, Italy. 
	 }}}
\address{
\centerline{Fermi National Accelerator Laboratory, Batavia, Illinois 60510}
}
%
% Indicate today's date
%
\date{\today}

\maketitle
%
% ==============> Text of the abstract goes here <=====================
% 
\begin{abstract}
We report on a 
%%%
preliminary
%%%
measurement of the ratio of the differential cross sections
for $W$ and $Z$ boson production as a function of transverse momentum 
in proton-antiproton collisions at {\mbox{$\sqrt{s}$ =\ 1.8\ TeV}}.
It has been shown that the ratio between $W$ and $Z$ observables
can be reliably calculated using perturbative QCD, even when the individual
observables are not. We present the first comparison between the measurement
of a $W$ and $Z$ observable ratio and a purely perturbative calculation.
The use of the ratio reduces both experimental and theoretical uncertainites
and might result in smaller overall uncertainties on the measured $W$ mass 
and width, compared to currently used methods at hadron colliders.
\end{abstract}
%\pacs{PACS numbers 14.65.Ha, 13.85.Qk, 13.85.Ni}

%\vskip 1cm

\newpage
\begin{center}
%\input{list_of_authors.tex}
% LIST_OF_AUTHORS.TEX                 5/18/01            
%
\small{                                                                      
%% names begin here                                                           
V.M.~Abazov,$^{23}$                                                           
B.~Abbott,$^{58}$                                                             
A.~Abdesselam,$^{11}$                                                         
M.~Abolins,$^{51}$                                                            
V.~Abramov,$^{26}$                                                            
B.S.~Acharya,$^{17}$                                                          
D.L.~Adams,$^{60}$                                                            
M.~Adams,$^{38}$                                                              
S.N.~Ahmed,$^{21}$                                                            
G.D.~Alexeev,$^{23}$
% ...
\\                                                                            
\vskip 0.30cm                                                                 
\centerline{(D\O\ Collaboration)}                                             
\vskip 0.30cm                                                                 
}                                                                             
%\address{                                                                     
\small{\it
\centerline{$^{1}$Universidad de Buenos Aires, Buenos Aires, Argentina}       
\centerline{$^{2}$LAFEX, Centro Brasileiro de Pesquisas F{\'\i}sicas,         
                  Rio de Janeiro, Brazil}                                     
\centerline{$^{3}$Universidade do Estado do Rio de Janeiro,                   
                  Rio de Janeiro, Brazil}                                     
\centerline{$^{4}$Institute of High Energy Physics, Beijing,                  
                  People's Republic of China}                                 
\centerline{$^{5}$Universidad de los Andes, Bogot\'{a}, Colombia}             
\centerline{$^{6}$Charles University, Center for Particle Physics,            
                  Prague, Czech Republic}                                     
\centerline{$^{7}$Institute of Physics, Academy of Sciences, Center           
                  for Particle Physics, Prague, Czech Republic}               
\centerline{$^{8}$Universidad San Francisco de Quito, Quito, Ecuador}         
\centerline{$^{9}$Institut des Sciences Nucl\'eaires, IN2P3-CNRS,             
                  Universite de Grenoble 1, Grenoble, France}                 
\centerline{$^{10}$CPPM, IN2P3-CNRS, Universit\'e de la M\'editerran\'ee,     
                  Marseille, France}                                          
\centerline{$^{11}$Laboratoire de l'Acc\'el\'erateur Lin\'eaire,              
                  IN2P3-CNRS, Orsay, France}                                  
\centerline{$^{12}$LPNHE, Universit\'es Paris VI and VII, IN2P3-CNRS,         
                  Paris, France}                                              
\centerline{$^{13}$DAPNIA/Service de Physique des Particules, CEA, Saclay,    
                  France}                                                     
\centerline{$^{14}$Universit{\"a}t Mainz, Institut f{\"u}r Physik,            
                  Mainz, Germany}                                             
\centerline{$^{15}$Panjab University, Chandigarh, India}                      
\centerline{$^{16}$Delhi University, Delhi, India}                            
\centerline{$^{17}$Tata Institute of Fundamental Research, Mumbai, India}     
\centerline{$^{18}$Seoul National University, Seoul, Korea}                   
\centerline{$^{19}$CINVESTAV, Mexico City, Mexico}                            
\centerline{$^{20}$FOM-Institute NIKHEF and University of                     
                  Amsterdam/NIKHEF, Amsterdam, The Netherlands}               
\centerline{$^{21}$University of Nijmegen/NIKHEF, Nijmegen, The               
                  Netherlands}                                                
\centerline{$^{22}$Institute of Nuclear Physics, Krak\'ow, Poland}            
\centerline{$^{23}$Joint Institute for Nuclear Research, Dubna, Russia}       
\centerline{$^{24}$Institute for Theoretical and Experimental Physics,        
                   Moscow, Russia}                                            
\centerline{$^{25}$Moscow State University, Moscow, Russia}                   
\centerline{$^{26}$Institute for High Energy Physics, Protvino, Russia}       
\centerline{$^{27}$Lancaster University, Lancaster, United Kingdom}           
\centerline{$^{28}$Imperial College, London, United Kingdom}                  
\centerline{$^{29}$University of Arizona, Tucson, Arizona 85721}              
\centerline{$^{30}$Lawrence Berkeley National Laboratory and University of    
                  California, Berkeley, California 94720}                     
\centerline{$^{31}$University of California, Davis, California 95616}         
\centerline{$^{32}$California State University, Fresno, California 93740}     
\centerline{$^{33}$University of California, Irvine, California 92697}        
\centerline{$^{34}$University of California, Riverside, California 92521}     
\centerline{$^{35}$Florida State University, Tallahassee, Florida 32306}      
\centerline{$^{36}$University of Hawaii, Honolulu, Hawaii 96822}              
\centerline{$^{37}$Fermi National Accelerator Laboratory, Batavia,            
                   Illinois 60510}                                            
\centerline{$^{38}$University of Illinois at Chicago, Chicago,                
                   Illinois 60607}                                            
\centerline{$^{39}$Northern Illinois University, DeKalb, Illinois 60115}      
\centerline{$^{40}$Northwestern University, Evanston, Illinois 60208}         
\centerline{$^{41}$Indiana University, Bloomington, Indiana 47405}            
\centerline{$^{42}$University of Notre Dame, Notre Dame, Indiana 46556}       
\centerline{$^{43}$Iowa State University, Ames, Iowa 50011}                   
\centerline{$^{44}$University of Kansas, Lawrence, Kansas 66045}              
\centerline{$^{45}$Kansas State University, Manhattan, Kansas 66506}          
\centerline{$^{46}$Louisiana Tech University, Ruston, Louisiana 71272}        
\centerline{$^{47}$University of Maryland, College Park, Maryland 20742}      
\centerline{$^{48}$Boston University, Boston, Massachusetts 02215}            
\centerline{$^{49}$Northeastern University, Boston, Massachusetts 02115}      
\centerline{$^{50}$University of Michigan, Ann Arbor, Michigan 48109}         
\centerline{$^{51}$Michigan State University, East Lansing, Michigan 48824}   
\centerline{$^{52}$University of Nebraska, Lincoln, Nebraska 68588}           
\centerline{$^{53}$Columbia University, New York, New York 10027}             
\centerline{$^{54}$University of Rochester, Rochester, New York 14627}        
\centerline{$^{55}$State University of New York, Stony Brook,                 
                   New York 11794}                                            
\centerline{$^{56}$Brookhaven National Laboratory, Upton, New York 11973}     
\centerline{$^{57}$Langston University, Langston, Oklahoma 73050}             
\centerline{$^{58}$University of Oklahoma, Norman, Oklahoma 73019}            
\centerline{$^{59}$Brown University, Providence, Rhode Island 02912}          
\centerline{$^{60}$University of Texas, Arlington, Texas 76019}               
\centerline{$^{61}$Texas A\&M University, College Station, Texas 77843}       
\centerline{$^{62}$Rice University, Houston, Texas 77005}                     
\centerline{$^{63}$University of Virginia, Charlottesville, Virginia 22901}   
\centerline{$^{64}$University of Washington, Seattle, Washington 98195}       
\vskip 0.5 cm
\centerline{$^{*}$ Visitor from University of Zurich, Zurich, Switzerland.}
}                                                                             
%end                                                                          

\end{center}

\normalsize

\vfill\eject

\section{Introduction}
The D\O\ Collaboration 
has recently published~\cite{wpt,zpt} measurements of the 
differential cross sections for $W$ and $Z$ boson production  
as a function of transverse momentum $p_T$. Both measurements are in good
agreement with combined resummed and perturbative QCD models, such
as those in references~\cite{ak,ly,ev}. For the analyses based on data taken
by D\O\ during 1992--1996 (Tevatron Run 1), we have 
used the resummed calculation of Ref.~\cite{ly} fitted to our own Drell--Yan 
data to determine the non-perturbative phenomenological parameters of the
theory. The resummed calculation was then used to predict 
$W$ boson observables such as the
electron and neutrino transverse momenta. These predictions are used as
input to a Monte
Carlo model of $W$ boson production and decay, which is used in our published
measurements of the $W$ boson mass~\cite{wmass} 
and $W$ boson production cross section~\cite{xsections}. 

Reference~\cite{gk} proposes an alternative method of predicting $W$ boson
observables from measured $Z$ boson quantities using the theoretical
ratio of the $W$ to $Z$ boson differential cross sections with respect
to variables that have been scaled by the corresponding vector boson masses.
Because $W$ and $Z$ boson production properties are very similar, 
the large radiative corrections that affect the individual distributions 
cancel when calculating the ratio. 
The ratio can therefore be reliably
calculated using perturbative QCD (pQCD), 
with no need for resummation, even for
small values of the transverse momenta of the vector bosons.
In fact, as the transverse momentum of the vector boson becomes smaller,
the radiative corrections factorize from the hard process and cancel when
taking the ratio. The theoretical uncertainties are therefore well
understood, and smallest, at very low $p_T$. 

The basic proposal 
is to combine pQCD calculations with the measured
$Z$ boson observables to extract the $W$ boson observables. This method
is exemplified and tested here
using the $W$ and $Z$ boson differential cross 
sections as a function of transverse momentum. The main 
difference between the $W$ and the $Z$ boson production properties 
is due to the
difference in the $W$ and $Z$ boson masses. 
We therefore work with vector boson variables that have been scaled by
the corresponding vector boson mass $m_V$.

The ratio of differential cross sections with respect to the scaled $W$  and
$Z$ boson transverse momenta $p_T^W/m_W$ and $p_T^Z/m_Z$ is defined as
\begin{equation}
R_{p_T}=\left(\frac{d\sigma^W}{dp_T^W/m_W}\right) {\Big /}
\left( \frac{d\sigma^Z}{dp_T^Z/m_Z} \right) ,
\label{eq:ratiodef}
\end{equation}
where $d\sigma^V/dp_T^V$ is the differential cross section for
vector boson production $\sigma(p\bar p \to V + X)$ with respect to 
the vector boson transverse momentum $p_T^V$.
Equation~\ref{eq:ratiodef} can be used to predict the differential
cross section for $W$ bosons with respect to the non-scaled 
transverse momentum:
\begin{equation}
\left. \frac{d\sigma^W}{dp_T^W}  \right|_{\rm predicted}=
\frac{m_Z}{m_W} \times R_{p_T} \times  
\left. \frac{d\sigma^Z}{dp_T^Z}  \right|_{\rm measured}^{p_T^Z=\frac{m_Z}{m_W} p_T^W} ,
\label{eq:wptdef}
\end{equation}
where $R_{p_T}$ is calculated using pQCD.
In this paper, we present the first measurement of $R_{p_T}$, and confront it
with the calculation of Ref.~\cite{gk}. 

Next, we use our measured differential $Z$ boson cross section,
Eq.~\ref{eq:wptdef}
and Ref.~\cite{gk} to obtain the differential 
$W$ boson cross section and compare it to our 
published result~\cite{wpt}.

Compared to the method previously used to extract $W$ boson observables, 
the present ratio method reduces
both theoretical and experimental systematic uncertainties. However,
it introduces a statistical contribution to the uncertainty
from the number of events in the $Z$ boson candidate sample.
Such a trade--off will eventually
result in smaller overall uncertainties, especially when used
with the high statistics samples expected during Run 2 at the Tevatron.

\section{Data Selection}
We keep the modifications to the published D\O\
analyses~\cite{wpt,zpt} to a minimum,
but, at the same time, we try to cancel as many experimental
uncertainties as possible when measuring the ratio $R_{p_T}$. The 
uncertainty in the integrated luminosity of the data samples ($4.3\%$) is the
dominant uncertainty in the individual cross section measurements.
It cancels completely when taking the ratio, as long as the same
data sets are used to select the $W$ and $Z$ boson candidate samples.
In this analysis, we keep the event selection and 
corrections for background, efficiency, acceptance and detector resolutions
identical to the ones in the published results~\cite{wpt,zpt}, 
but require that the same data--taking runs be included in the 
$W$ and $Z$ boson event samples. In addition, we exclude events for which
large beam losses from the Main Ring accelerator were 
expected~\cite{xsections}. These beam losses can create significant
energy deposits in the calorimeter, resulting in events with large false 
transverse momentum imbalance, that could pass our $W$ selection cuts.
Due to these additional requirements, 
the $Z$ boson sample is reduced from 6407 to 4881 events.
About half of the event losses are due to the tightening of the
beam quality conditions, while the other half are events taken during runs 
for which there was no $W$ trigger present or it was prescaled.
The $W$ sample is reduced from 50488 to 50264 events when we remove
runs for which the $W$ trigger was prescaled.
The integrated luminosity for both samples is $(84.5 \pm 3.6)\rm pb^{-1}$. 

We have investigated whether 
additional sources of error could be cancelled in the
ratio. There are four sources of systematic error that contribute 
to the $W$ and $Z$ boson cross sections.
These arise from uncertainties in the background estimate, the event selection 
efficiency and the unfolding procedure used to correct for acceptance and 
detector resolution.

The dominant sources of background in both the
$W$ and the $Z$ boson analyses come from multijet and photon--jet events, where
the jets pass our electron identification criteria. In the case of the $W$,
a large imbalance in the transverse energy has to arise to fake the presence
of a neutrino. The mechanisms for multijet or photon--jet events 
to fake a $W$ or a $Z$ boson event
are different, and the methods used to estimate the background are completely
independent. We therefore cannot cancel any contribution to the error in the
ratio from the background estimates.

The acceptance and unfolding corrections are applied 
together using a parametric
Monte Carlo~\cite{wmass}. The main contribution to the error comes from the
detector modeling. For the $W$ analysis, we rely on the measurement of the 
energy of the recoiling hadrons, whereas the $Z$ boson measurement uses the 
measurement of the electromagnetic energy
from the electrons. We therefore do not attempt any cancellation of errors
in the acceptance/unfolding procedure either.

The uncertainty in the 
efficiency has contributions from the trigger and offline
electron identification. The Level 0 trigger, 
which requires the detection
of an inelastic collision via simultaneous hits in the forward and
backward Level 0 scintillation detectors~\cite{d0detector}, 
is common for $W$ and $Z$ boson events.
The uncertainty in this trigger  can therefore be completely cancelled
in the ratio. However, its contribution to the error on the
efficiency is negligible (0.5\% out of a total of 3.5\%). 

Although the triggers and the offline electron identification 
used in the $W$ and $Z$ boson analyses are different, 
the main contribution to the error in the efficiency (3\%) comes from
a common source, the so-called $u_{||}$ efficiency~\cite{wmass}.
This inefficiency arises
when the energy flow close to the electron increases as the recoiling
hadrons
approach the electron. It is therefore a topological effect produced
by the proximity of the electron to the jet, which has the largest effect
at a boson transverse momentum of about $20\;\rm GeV$~\cite{zpt}.
The parameterization of the  $u_{||}$ efficiency 
is done on an electron--by--electron
basis in the parametric Monte Carlo.
The error on the $u_{||}$ efficiency is estimated from Herwig~\cite{herwig}
$W$ and $Z$ boson
events, overlayed with data taken from randomly selected
{\mbox{$p\bar p$}}\ collisions. Because this
inefficiency depends on the proximity between electrons and jets 
in the event, it is difficult to estimate {\it a priori} 
how much of the uncertainty in the
$u_{||}$ efficiency cancels in the ratio. 
To determine if further investigations of the possible
cancellation of the $u_{||}$ efficiency error are worthwhile, we estimated 
the effect on the $R_{p_{T}}$
measurement of cancelling the contribution from the
$u_{||}$ efficiency error completely. This resulted in a maximum
reduction on the uncertainty on the ratio of less than 5\%. 
We therefore conclude that no error
cancellations beyond the errors in the luminosity would improve
the ratio measurement significantly.

\section{Scaled $W$ and $Z$ Boson Cross Sections}

Equation~(\ref{eq:ratiodef}) can be written
\begin{equation}
R_{p_T}^{th}=\left(\frac{d\sigma^W}{dp_T^W}\right) {\Big /}
\left( \frac{d\sigma^Z}{d(p_T^Z \times m_W/m_Z)} \right) 
= \left(\frac{d\sigma^W}{dp_T^W}\right) {\Big /}
\left( \frac{d\sigma^Z}{dp_T^{Z \prime}} \right) ,
\label{eq:ratiodef2}
\end{equation}

\noindent
where we have defined
\begin{equation}
p_T^{Z \prime} =  \frac{m_W}{m_Z} p_T^Z .
\label{eq:zptprime}
\end{equation}

\noindent
We use the value from the Particle Data Book~\cite{pdb} for the mass ratio
\begin{equation}
\frac{m_W}{m_Z}=0.8820\pm0.0005 .
\label{eq:massratio}
\end{equation}

\noindent
In order to measure the scaled distributions without changing the
binning of both the $W$ and the $Z$ 
boson analyses, we kept the $W$ bin boundaries
($\delta_i$) identical to the ones used in the published result. Since
we require the same bin widths for the scaled variables $p_T^W/m_W$
and $p_T^Z/m_Z$, we set the bin boundaries in the differential $Z$ boson 
cross section to
\begin{equation}
\delta_i \times \frac{m_Z}{m_W} = \delta_i/0.8820,
\end{equation}
and recompute the differential $Z$ boson cross section. 

Table~\ref{tab:d0data} shows the modified results for the
$W$ and $Z$ boson cross sections. The statistical
and systematic contributions to the individual
cross section uncertainties are shown separately. One can see that
the error on the ratio will be dominated by the systematic error
on the $W$ cross section.

%\clearpage
\section{Measurement of $R_{p_T}$}

\begin{figure}[h]
\centerline{\psfig{figure=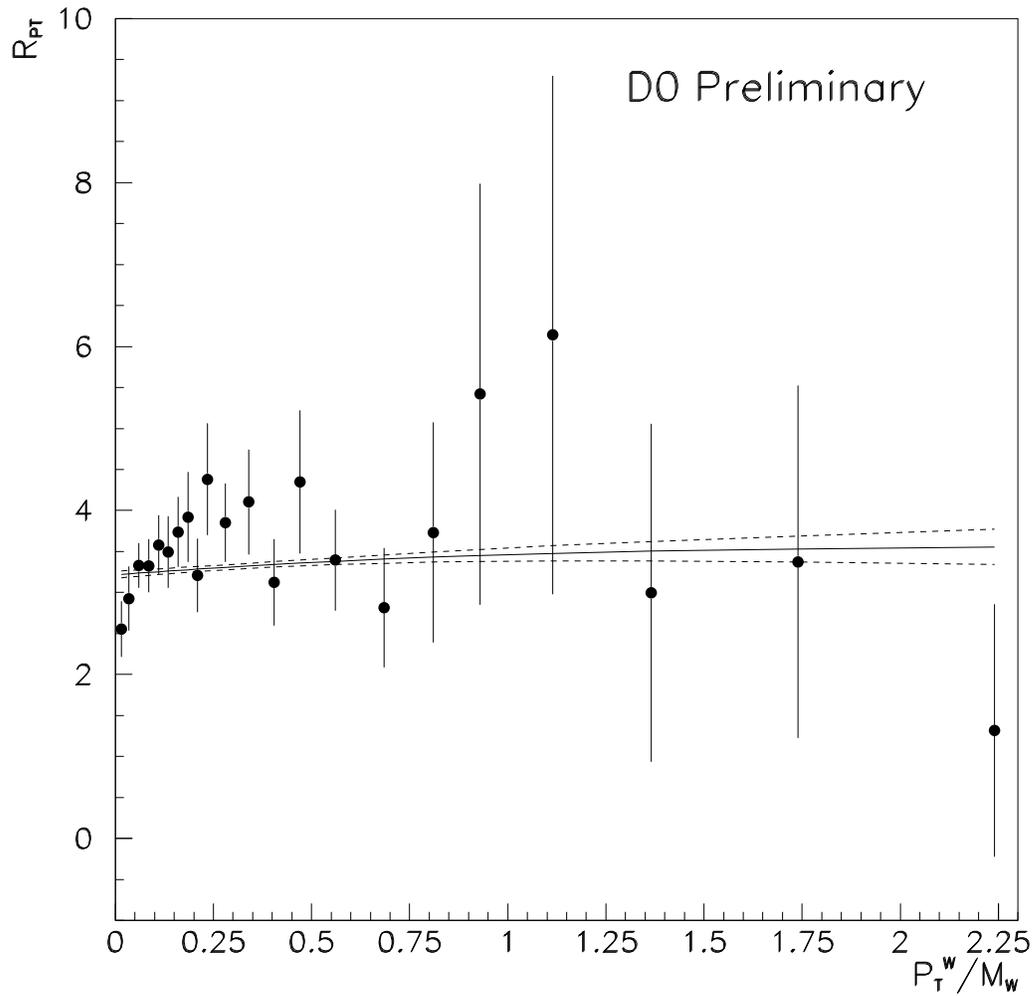,width=15cm}}
\caption{Ratio of scaled differential cross sections for $W$ to $Z$ production.
The solid line is the order $\alpha_S^2$ theoretical prediction of
Ref.~\protect\cite{gk} and  the dotted lines are the one sigma uncertainties
due to Monte Carlo integration.
The error on the luminosity cancels completely on the ratio of the
measured cross sections. No further error cancellation is attempted.}
\label{fig:rpt}
\end{figure}

Based on the measured $W$ and $Z$ boson differential cross sections listed in 
Table~\ref{tab:d0data}, we extract the ratio of scaled cross sections
\begin{equation}
R_{p_T}^{exp}=\left[ \left(\frac{d\sigma^W}{dp_T^W}\right) {\Big /}
\left( \frac{d\sigma^Z}{dp_T^Z} \right)\right] \times
\frac{m_W}{m_Z}\times \frac{Br(Z\to ee)}{Br(W\to e\nu)} .
\label{eq:dataratio}
\end{equation}
Note that the prediction for $R_{p_T}$~\cite{gk} was calculated for the
ratio of the scaled $W$ and $Z$ boson differential cross sections
$d\sigma^V/dp_T^V$, and we
measure the differential cross sections times their branching fractions to
electrons $(d\sigma^V/dp_T^V)\times Br(V \to e)$. We therefore
need to multiply our measurement by the ratio of the
$Z$ to $W$ boson branching fractions, which we obtain from the 
particle data book~\cite{pdb}:
\begin{equation}
Br(Z\to ee)=0.03367 \pm 0.00005
\end{equation}
\begin{equation}
Br(W\to e\nu)=0.1066 \pm 0.0020 .
\end{equation}
The result is shown in Figure~\ref{fig:rpt} and summarized in 
Table~\ref{tab:ratio}. 
The data are plotted at the value of
$p_T$ for which the theoretical prediction for $R_{p_T}$ is equal to its
average over the bin, following the prescription of reference~\cite{xbins}. 
We observe that the measured differential cross section
ratio agrees well with the purely pQCD theoretical 
prediction~\cite{gk},
even at small values of $p_T$. The $\chi ^2$ test for the comparison
between data and theory is $18.7$ for 21 degrees of freedom.

\section{Extraction of $\dsdpt$}
\begin{figure}[h]
\centerline{\psfig{figure=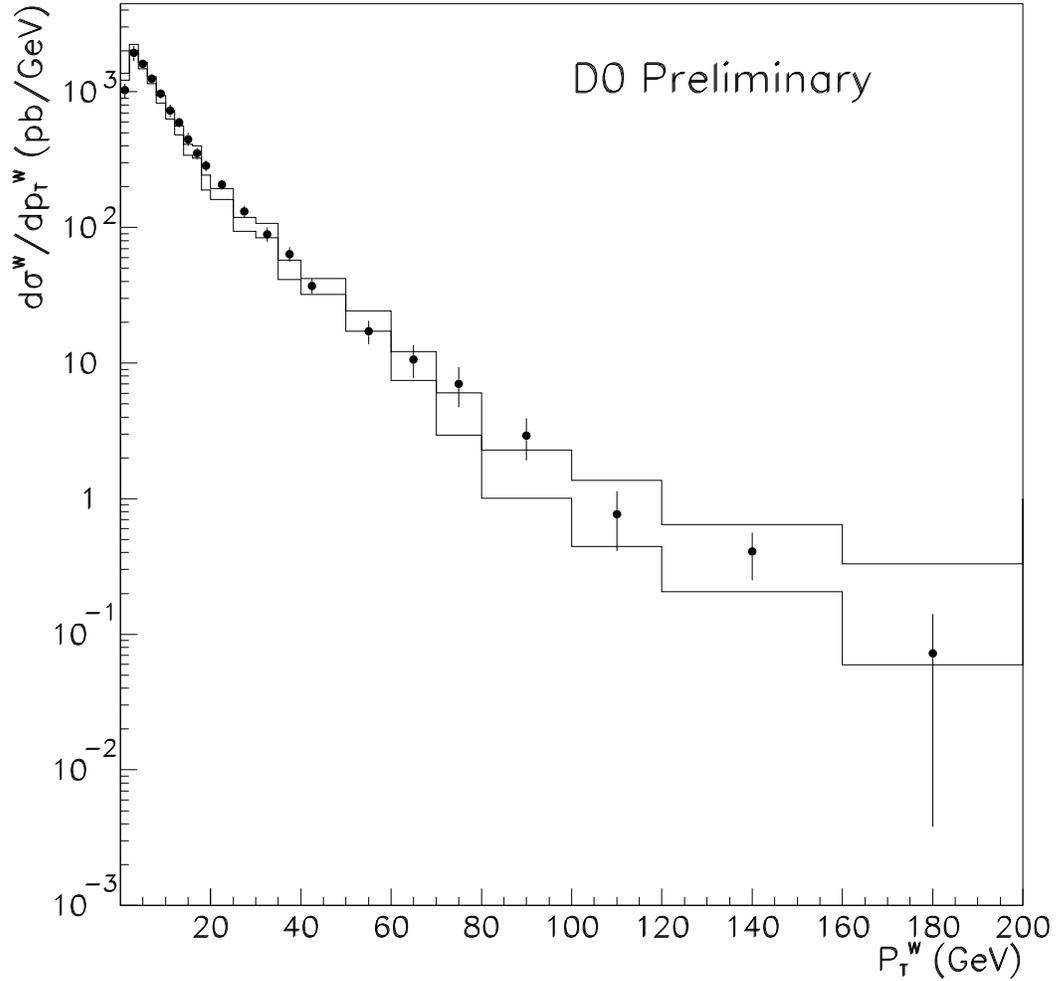,width=15cm}}
\caption{Differential cross section for $W$ boson production as a function
of $p_T^W$. The points are the D\O\ data; the error bars do not include
the $4.3\%$ error in the luminosity. The histograms represent the 
upper and lower 68\%
confidence level limits of the prediction~\protect\cite{gk} 
obtained from the ratio method.}
\label{fig:wpt}
\end{figure}

Based on Eq.~\ref{eq:wptdef}, we 
use the calculated ratio $R_{p_T}$ from reference~\cite{gk}, 
together with the measured 
$d\sigma^Z/dp_T^Z$, to predict the $W$ boson transverse
momentum spectrum and
compare it with our previously measured $d\sigma^W/dp_T^W$~\cite{wpt}. 
The result is
shown in Figure~\ref{fig:wpt}. The measured differential cross section
is plotted at the center of the bin. 
The ratio method prediction upper and lower 68\% confidence level 
limits are plotted as histograms.
The extracted transverse momentum distribution agrees very well with the
measurement; the Kolmogorov--Smirnov probability~\cite{kstest} 
$\kappa$ is equal to 1.

\section{Conclusions}
We have measured the ratio of the scaled differential cross sections
for $W$ to $Z$ boson production for the first time, and compared it to a purely
pQCD prediction. 
Our preliminary measurement shows good agreement with theory
%%
%We observe good agreement between data and theory
over the whole $p_T$ spectrum, even at small values of $p_T$.

We have used the theoretical prediction
for $R_{p_T}$, together with our measurement of the differential $Z$ boson 
production cross section, to obtain the $W$ differential cross section. 
This prediction agrees well with our published result. 

This method for
predicting $W$ properties should eventually
result in smaller overall uncertainties 
on the measured $W$ mass 
and width, compared to currently used methods at hadron colliders, once
the high statistics $Z$ boson samples expected during Run 2 at the Tevatron
become available.

\section*{Acknowledgements}
\label{sec:ack}
%
% =========> Choose the first acknowledgements paragraph <==========
% =========> if you are writing a physics paper. <==================
%
We thank Stephane Keller for interesting discussions
leading to this measurement and Walter Giele for
providing the theoretical prediction and assistance with technical
details to complete this measurement.
% Acknowledgement_paragraph.tex
%
We thank the staffs at Fermilab and collaborating institutions, 
and acknowledge support from the 
Department of Energy and National Science Foundation (USA),  
Commissariat  \` a L'Energie Atomique and 
CNRS/Institut National de Physique Nucl\'eaire et 
de Physique des Particules (France), 
Ministry for Science and Technology and Ministry for Atomic 
   Energy (Russia),
CAPES and CNPq (Brazil),
Departments of Atomic Energy and Science and Education (India),
Colciencias (Colombia),
CONACyT (Mexico),
Ministry of Education and KOSEF (Korea),
CONICET and UBACyT (Argentina),
The Foundation for Fundamental Research on Matter (The Netherlands),
PPARC (United Kingdom),
Ministry of Education (Czech Republic),
and the A.P.~Sloan Foundation.
%

%LIST_OF_VISITOR_ADDRESSES.TEX                            
%\input{list_of_visitor_addresses.tex}
%%%%%

\clearpage
\begin{table}[t]
\caption{Summary of the measured differential $W$ 
and $Z$ bosons production cross sections as a function
of transverse momentum used to calculate the ratio. 
The error on the ratio is dominated by the systematic error
on the $W$ cross section.}
\begin{center}
\begin{tabular}{||c|c|c|c|c|c|c||}
Bins & {\mbox{$\frac{d\sigma(W\to e \nu)}{dp_T^{W}}$}}&
Stat Error & Syst Error & 
{\mbox{$\frac{d\sigma(Z\to e^+ e^-)}{dp_T^Z}$}}& Stat Error
& Syst Error\\
\scriptsize{(GeV)} & \scriptsize{(pb/GeV)} & \scriptsize{(pb/GeV)} &
\scriptsize{(pb/GeV)} & \scriptsize{(pb/GeV)}
& \scriptsize{(pb/GeV)} & \scriptsize{(pb/GeV)} \\
\hline \hline %$
 0-2 &  109.48 &  4.61   & 12.35 &      11.94 & 0.53 &  0.35 \\ \hline 
 2-4 &  206.21 & 6.85   & 24.64  &      19.63 & 0.65 &  0.57 \\ \hline
 4-6 &  171.32 & 5.65   & 9.29 &      14.34 & 0.53 & 0.44 \\ \hline
 6-8 & 133.60 &  4.65   & 9.46 &      11.19 & 0.48 &  0.36 \\ \hline
 8-10 & 103.48 &  4.04   & 6.95  &      8.05 & 0.41 &   0.27 \\ \hline
10-12 & 77.46 &  3.46   & 7.25 &     6.18 &    0.37 &  0.21 \\ \hline 
12-14 &  63.58 & 3.20   & 4.16  &    4.74 &   0.33 &   0.15 \\ \hline  
14-16 &   47.77&  2.77  & 4.29 &      3.39 &   0.28 &  0.11 \\ \hline
16-18 &  37.67 &  2.42     & 2.73 &       3.27 & 0.28 &   0.17 \\ \hline
18-20 & 30.50 &    2.20   & 1.74 &       1.94 & 0.22 &  0.11 \\ \hline
20-25 & 22.02 &    1.23   & 1.22 &      1.59&  0.12 &   0.08  \\ \hline
25-30 & 13.94&    0.93   &   1.07 &     0.946 & 0.097 &  0.051 \\ \hline
30-35 &  9.51 &   0.73   &  0.84 &     0.848 &  0.092 &   0.043 \\ \hline
35-40 &  6.79&    0.63   &  0.51&     0.435&   0.066 &  0.022 \\ \hline
40-50 & 3.96 &    0.37   &  0.31 &     0.325 & 0.040 &   0.016 \\ \hline
50-60 & 1.82 &   0.25   &  0.25 &     0.180 &  0.029 &  0.009  \\ \hline
60-70 & 1.14 &    0.20   &  0.23 &    0.0848 & 0.0197 &   0.0045 \\ \hline
70-80 &  0.749 &   0.178  &   0.170   &     0.0385 & 0.0129 & 0.0020 \\ \hline
80-100 &  0.310&   0.059 & 0.088 &     0.0141 &  0.0054 &   0.0008 \\ \hline
100-120 &  0.0822&  0.0287& 0.0255 & 0.00764& 0.00383 &  0.00032 \\ \hline
120-160 &  0.0433& 0.0119& 0.0118& 0.00358&  0.00180 &  0.00018 \\ \hline
160-200 &  0.00769& 0.00545 & 0.00482& 0.00163& 0.00111 &  0.00010 \\
\end{tabular}
\end{center}
\label{tab:d0data}
\end{table}

\clearpage

\begin{table}[t]
\caption{Measured $R_{p_T}$. The uncertainty in the luminosity
of the $W$ and $Z$ samples cancels completely when taking the
ratio. No further error cancellation is attempted.}
\begin{center}
\begin{tabular}{ ||c|c|c|c||}
Bins \scriptsize{(GeV)} & $p_T$ \scriptsize{(GeV)} & $R_{p_T}$ & Error \\ 
\hline \hline %$
 0-2 & 1.21 &  2.555  &   0.340 \\ \hline 
 2-4 &  2.81 & 2.927  &   0.388  \\ \hline
 4-6 &  4.83 &  3.327 &     0.273  \\ \hline
 6-8 &  6.84 &  3.326 &     0.323 \\ \hline
 8-10 & 8.85 &   3.580 &     0.360 \\ \hline
10-12 & 10.86 &   3.494 &     0.439\\ \hline 
12-14 & 12.87 &   3.739 &     0.426 \\ \hline  
14-16 & 14.88 &   3.921 &     0.550 \\ \hline
16-18 & 16.89 &  3.208 &     0.450 \\ \hline
18-20 & 18.90 &   4.380 &     0.683\\ \hline
20-25 & 22.52 & 3.854 &    0.479 \\ \hline
25-30 & 27.34 & 4.105 &     0.640 \\ \hline
30-35 & 32.57 & 3.125 &     0.529 \\ \hline
35-40 & 37.89 &  4.348 &     0.875 \\ \hline
40-50 & 45.03 &  3.395 &     0.615  \\ \hline
50-60 & 55.09 & 2.814 &     0.727\\ \hline
60-70 & 65.14 &  3.731 &      1.342\\ \hline
70-80 & 74.79 &  5.419 &      2.566 \\ \hline
80-100 & 89.67 &  6.140 &      3.160 \\ \hline
100-120 & 109.77 & 2.995 &      2.060 \\ \hline
120-160 & 139.93 & 3.374 &      2.151  \\ \hline
160-200 & 180.14 & 1.317 &      1.539 \\ 
\end{tabular}
\end{center}
\label{tab:ratio}
\end{table}
\clearpage

\clearpage

\clearpage
\end{document}